 \documentclass[preprint2]{aastex}

\shorttitle{Reinstating RX\thinspace J0042.3+4115 as a black hole binary}
\shortauthors{Barnard et al.}

\begin{document}


\title{Reinstating the M31 X-ray system RX J0042.3+4115 as a black hole X-ray binary, and compelling evidence for an extended corona }


\author{R. Barnard, and  M. R.  Garcia}
\affil{Harvard-Smithsonian Center for Astrophysics, 60 Garden St, Cambridge MA 02138}
\and
\author{S. S. Murray}
\affil{Johns Hopkins University, Baltimore, Maryland}


\begin{abstract}
The M31 X-ray source RX\thinspace J0042.3+4115 was originally identified as a black hole binary because it  displayed characteristic low state variability at conspicuously high luminosities; unfortunately, this variability was later found to be artificial. However, analysis of 84 Chandra ACIS observations, an HST ACS/WFC observation, and a 60 ks XMM-Newton observation has supplied new evidence that RX\thinspace  J0042.3+4115 is indeed a black hole binary.  The brightest optical star within  3$\sigma$ of the position of RXJ0042.3+4115 had a F435W ($\sim$B) magnitude of 25.4$\pm$0.2; $M_{\rm B}$ $>$ $-$0.4, hence we find a low mass donor likely.
RX J0042.3+4115 was persistently bright over $\sim$12 years. Spectral fits revealed  characteristic black hole binary states: a low/hard state at 2.08$\pm$0.08$\times 10^{38}$ erg s$^{-1}$, and a steep power law state at 2.41$\pm$0.05$\times 10^{38}$ erg s$^{-1}$ (0.3--10 keV). The high luminosity low state suggests a $\sim$20 $M_{\odot}$ primary; this is  high, but within the range of known stellar black hole masses. The inner disk temperature during the steep power law state is 2.24$\pm$0.15 keV, high but strikingly similar to that of GRS\thinspace 1915+105, the only known Galactic black hole binary with a low mass donor to be persistently bright.  Therefore RX\thinspace  J0042.3+4115 may be an analog for GRS\thinspace 1915+105; however, other mechanisms may account for its behavior.  We  find  compelling evidence for an extended corona during the steep power law state, because compact corona models where the seed photons for Comptonization are tied to the inner disc temperature are rejected.
\end{abstract}





\section{Introduction}

The M31 X-ray source RX\thinspace J0042.3+4115, named following \citet{supper97}, was originally identified as a black hole X-ray binary by \citet{barnard03}, after analyzing 4 XMM-Newton observations from 2000--2002. It apparently exhibited power density spectra  (PDS) that were well described by a broken power law, with spectral index $\alpha$ changing from $\sim$0 to $\sim$1 at some break frequency;  such a  PDS is  consistent with the low/hard states observed in all X-ray binaries, whether the accretor is a neutron star or black hole \citep[see e.g.][]{vdk94,wijnands99}. Neutron star X-ray binaries tend to exhibit such behaviour at luminosities around 10$^{36}$--10$^{37}$ erg s$^{-1}$, yet RX\thinspace J0042.3+4115 exhibited this variability at 0.3--10 keV luminosities of $\sim$1--3$\times 10^{38}$ erg s$^{-1}$. \citet{barnard03} concluded that RX\thinspace J0042.3+4115 contained a black hole.

However, it was later discovered that these PDS  and those reported by other groups were contaminated by artifacts caused by  the  XMM-Newton data reduction software \citep{barnard07}. The problem arose because all XMM-newton lightcurves start at the arrival time of the first photon by default; hence, source and background lightcurves, and lightcurves from the three EPIC detectors--- MOS1, MOS2, and pn--- are asynchronous by default. Combining these lightcurves (e.g. combining instruments, or background subtraction)  often resulted in PDS  with artificial broken power law shapes.

Therefore we make no use of the PDS and instead rely on our well established method of using  low state emission spectra \citep[power law emission with photon index 1.4--1.7 and little to no thermal emission, ][]{mr06} at conspicuously high luminosities to identify black hole candidates  \citep{barnard08, barnard09,barnard11b}. We present the most  detailed justification of our selection criteria in \citet{barnard11b}.

In this paper  we present our analysis of 84 Chandra ACIS observations of RX\thinspace J0042.3+4115 over $\sim$12 years, and our serendipitous HST observation,  along with our re-analysis of the 60 ks 2002 XMM-Newton observation. We use the HST data to place RX\thinspace J0042.3+4115 in M31, and argue for a low mass donor. We use long-term and short-term variability, and also emission spectra, to reinstate RX\thinspace J0042.3+4115 as a black hole candidate. We discuss the observations and data analysis in the next section, followed by our results in Section 3, and a discussion in Section 4.

\section{Observations and analysis}

We analyzed 84 Chandra ACIS observations of the central region of M31, spaced over $\sim$12 years, using CIAO version 4.3. For each observation we extracted  0.3--7.0 keV source and background spectra from circular regions with 10$''$ radius; the background region was close to the source region, and source free. Corresponding response matrices and ancillary response files were also made. We obtained 0.3--10 keV luminosities from each observation using XSPEC version 12.6.0. 

Observations with $>$200 net source counts were freely fitted with absorbed power law models; spectra were grouped to give at least 20 counts per bin. For  observations with $<$200 net source counts we assumed an absorbed power law model with $N_{\rm H}$ =1.0$\times$10$^{21}$ atom cm$^{-2}$ and $\Gamma$ = 1.5, and found the 0.3--10 keV luminosity equivalent to 1 count s$^{-1}$, then multiplied this conversion factor by the intensity; we chose this model because it approximates the best fit to our deepest Chandra observation of RX\thinspace J0042.3+4115 in its low state. Luminosity uncertainties for freely fitted spectra are estimated by XSPEC by calculating a range of fluxes obtained by varying the emission parameters; the uncertainties for the faint spectra are derived directly from intensity uncertainties.

Additionally, we analyzed the 60 ks 2002 January XMM-Newton observation of M31 (Rev 381) with SAS version 10.0.0. We extracted  0.3--10 keV EPIC-pn lightcurves and spectra from circular  source and background regions with 15$''$ radius. The background region was on the same CCD as the source region,  and at a similar off-axis angle. The spectra were grouped to ensure a minimum of 50 counts per bin.

RX\thinspace J0042.3+4115 was serendipitously observed in one of our  HST ACS/WFC observations of M31 transients. Observation j9ud17010 was made on 2009 August 25, with the F435W filter for 4360 s. We registered a combined Chandra 0.3--7.0 keV image and the HST image to the same B band image provided by the Local Group Galaxy Survey \citep{massey06} using the IRAF task {\sc ccmap}; we used X-ray bright globular clusters to register the Chandra image, and bright, unsaturated  stars to register the HST image. We determined the best X-ray position of RX\thinspace J0042.3+4115 with the IRAF task {\sc imcentroid}.  We used the IRAF package DAOPHOTII to perform photometry on the stars within 3$\sigma$ of the position of RX\thinspace J0042.3+4115.

\section{Results}

\subsection{The search for an optical counterpart}

The centroid of the X-ray emission from RX\thinspace J0042.3+4115 was located at RA = 00:42:22.954, Dec = 41:15:35.23, with 1$\sigma$ uncertainties of 0.009$''$ in RA and 0.007$''$ in Dec. Combining this with the r.m.s. uncertainties in registration yields X-ray positional uncertainties of 0.09$''$ in RA and 0.19$''$ in Dec. 

\begin{figure}
\epsscale{.8}
\plotone{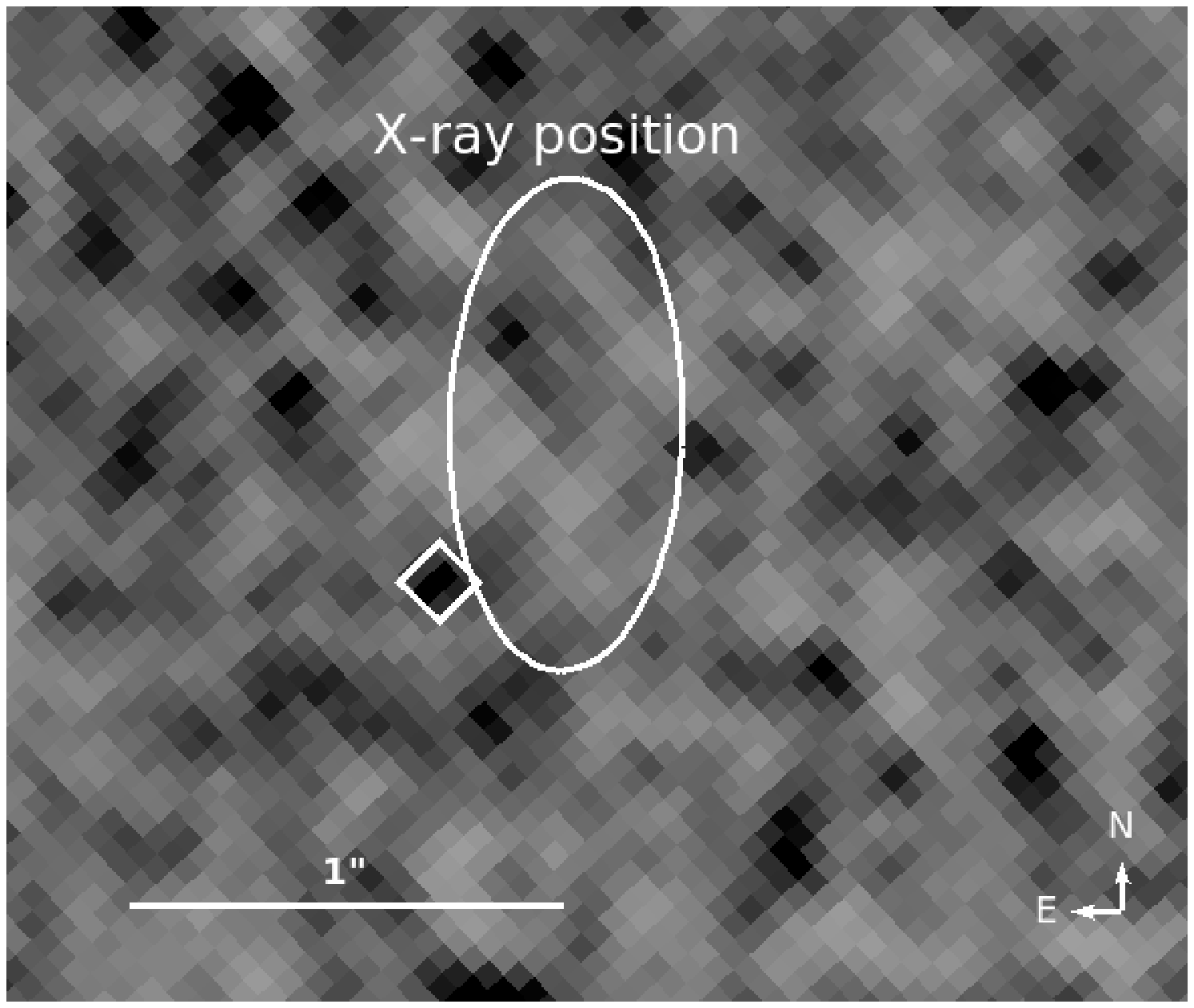}
\caption{A  detail of the HST ACS/WFC image from  observation j9ud17010; the exposure time was 4360 s, and  the F435W ($\sim$B band) filter was used. The white ellipse represents the 3$\sigma$ uncertainty in the position of RX\thinspace J0042.3+4115. North is up, east is left. The brightest star consistent with the ellipse is indicated by a diamond; it has a Vega B band  magnitude of 25.4$\pm$0.2. }\label{opt}
\end{figure}

 Figure~\ref{opt} shows a detail our HST image, superposed with an ellipse representing the 3$\sigma$ uncertainties in the position of RX\thinspace J0042.3+4115. There are several stars within the ellipse, the brightest of which has a Vega  B magnitude of 25.4$\pm$0.2. We therefore constrain the B band magnitude to $\ga$24.8. We see no evidence for a background galaxy; the region is relatively uncrowded, as it is $\sim$4$'$ from the M31 bulge, and the detection limit is B $<$ 28. There is no evidence for a  counterpart in the 2MASS All Sky Catalog \citep{skrutskie2006}, hence RX\thinspace J0042.3+4115 cannot be associated with a late type star in our Galaxy. We conclude that it is located in M31.

 The distance modulus for M31 $\sim$24.5, and we can estimate $A_{\rm B}$ by using the empirical relations $A_{\rm V}$ $\sim$ $N_{\rm H}$/1.8$\times 10^{21}$ atom cm $^{-2}$ \citep{predehl95}, and E(B-V) $\sim$ $A_{\rm V}$/3. The  column density varied significantly between observations, hence the variable component was probably internal to the system. Since the donor is unlikely to suffer this extra absorption, we assume  $N_{\rm H}$ = 1.0$\times$10$^{21}$ atom cm$^{-2}$ (see below), and $A_{\rm B}$ = 0.7. Therefore $M_{\rm B}$ $\ga$ $-$0.4.

The known counterparts of high mass X-ray binaries (HMXBs) in the SMC have apparent V magnitudes in the range 13 $\la$ $m_{\rm V}$ $\la$ 18, and $B-V$ in the range $-0.32$  $\le$ B-V $\le$  0.06 \citep[see e.g.][]{coe05,antoniou09}. For a distance of $\sim$60 kpc, this equates to $-$6 $\la$ $M_{\rm B}$ $\la$ $-1$, all brighter than our threshold of $M_{\rm B}$ $\ga$ $-$0.4. 

The three known BH HMXBs are Cygnus X-1, LMC X-1 and LMC X-3. Cygnus X-1 has a counterpart with $M_{\rm V}$ = $-$6.5 \citep{walborn72} and B$-$V = 0.8 \citep{hiltner56}. The counterpart to LMC X-1 has magnitude V = 14.60$\pm$0.02 and B$-$V = 0.17$\pm$0.08 \citep{orosz09}; hence $M_{\rm B}$ $\sim$ $-$4.2 for a distance of 50 kpc. LMC X-3 has a B $\sim$ 17 counterpart \citep[see e.g.][]{brocksopp01}, and $M_{\rm B}$ $\sim$ $-$1.5; this is $\sim$8$\sigma$ brighter than the brightest star within the  ellipse.

 RX\thinspace J0042.3+4115 is therefore most likely to be a low mass X-ray binary (LMXB).

\subsection{Time variability}

\begin{figure}
\epsscale{1}
\plotone{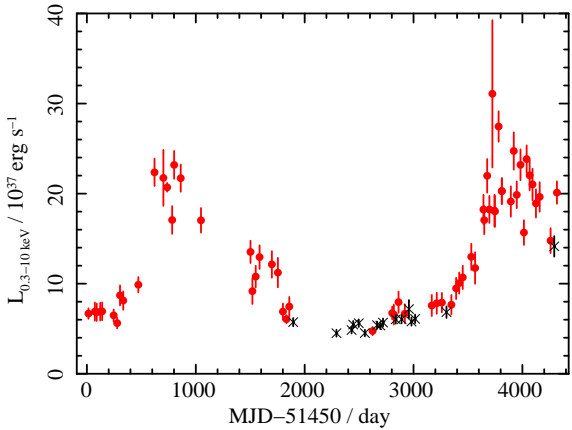}
\caption{Long-term, calibrated luminosity lightcurve of RX\thinspace J0042.3+4115 from $\sim$12 years of Chandra ACIS observations. The luminosity varied by a factor $\sim$6. Circles represent luminosities from freely fitted spectra, while crosses represent ``faint'' spectra. Uncertainties are quoted at the 1$\sigma$ level. } \label{acislc}
\end{figure}

We present the $\sim$12 year 0.3--10 keV luminosity lightcurve of RX\thinspace J0042.3+4115  created from the 84 ACIS observations in Fig.~\ref{acislc}; circles mark luminosities from freely-fitted bright spectra, while crosses represent faint spectra where we assume $N_{\rm H}$ = 1.0$\times$10$^{21}$ atom cm s$^{-2}$, and $\Gamma$ = 1.5. Uncertainties are quoted at a 1$\sigma$ level.  The lightcurve is extremely variable, with the luminosity varying over $\sim$0.5--3$\times$10$^{38}$ erg s$^{-1}$. We note that the spectral fits to all bright observations are consistent with a constant $\Gamma$, although $N_{\rm H}$ varied by a factor $\sim$5; the faint observations may be up to 40\% times brighter.

 RX\thinspace J0042.3+4115 appears to be persistently bright. By contrast,  most Galactic black hole LMXBS are transient. One exception is GRS 1915+105; a 7 year RXTE/ASM lightcurve of GRS\thinspace 1915+105 showed it to be persistently bright \citep{mr06}. Another possible exception is GRS 1758-258; it is thought to be a Galactic LMXB, but its true nature is not confirmed due to the high degree of absorption \citep[see e.g.][ and references within]{munoz_arjonilla10}

\begin{figure}
\epsscale{1}
\plotone{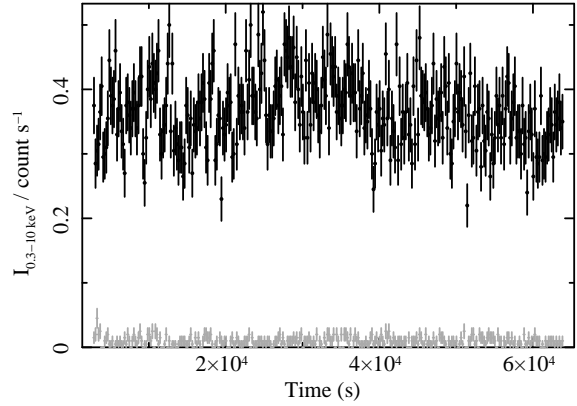}
\caption{EPIC-pn 0.3--10 keV lightcurve of RX\thinspace J0042.3+4115 from the 60 ks  Rev 381 observation. The background lightcurve is shown in grey for comparison. We see that RX\thinspace J0042.3+4115 is significantly variable over short time-scales, varying by a factor 2 in $<$10 ks. } \label{xmmlc}
\end{figure}

We also examined the short term variability of RX\thinspace J0042.3+4115 during the long XMM-Newton observation. We present the 0.3--10 keV EPIC-pn intensity lightcurve for RX\thinspace J0042.3+4115 in Fig.~\ref{xmmlc}, along with the background lightcurve in grey for comparison.  The intensity varies by a factor 2 (4$\sigma$ deviation) in $<$10 ks, hence the emission is dominated by a single source. The probability  of RX\thinspace J0042.3+4115 consisting of multiple bright variable sources is very low, especially since it is not associated with any globular cluster.

\subsection{Spectral analysis}
\subsubsection{ Chandra observation OBSID1575}

The longest ACIS observation of RX\thinspace J0042.3+4115 was OBSID1575, an ACIS-S observation  with a $\sim$40 ks exposure time; the net source spectrum contained 7690  photons. An on-axis source with this intensity would be in danger of pile-up; however,  RX\thinspace J0042.3+4115 was $\sim$4$'$ off-axis, and the photons  were spread over a large number of pixels ($>$100). Each incoming photon is assessed by its impact on  a 3x3 array of ACIS pixels ; ``good'' photons are detected in only 2 of the 9 pixels, while cosmic rays etc. are detected in 3 or more \citep{davis01}.  We therefore estimated the probability of pile up from the brightest pair of pixels; this pair accumulated 807 photons over $\sim$38 ks, or one photon every $\sim$14 frames. Hence, we conclude that pile up is unlikely to have been  significant.

\begin{figure}
\epsscale{1}
\plotone{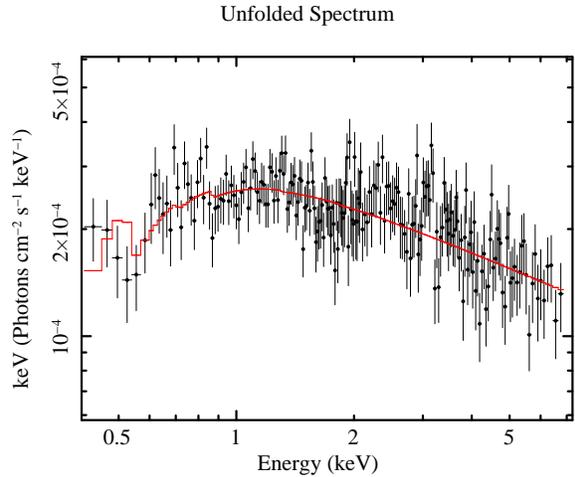}
\caption{Unfolded spectrum for ACIS observation 1575, multiplied by energy. It is well described by an absorbed power law with $N_{\rm H}$ = 1.0$\pm$0.2$\times 10^{21}$ atom cm$^{-2}$, and $\Gamma$ = 1.46$\pm$0.06; $\chi^2$/dof = 211/204. Such a spectrum is characteristic of the low/hard state seen in neutron star and black hole binaries \citep{mr06}.}\label{1575spec}
\end{figure}

The 0.3--7.0 keV spectrum of RX\thinspace J0042.3+4115 during observation OBSID1575 is well described by an absorbed power law, with line-of-sight absorption  $N_{\rm H}$ = 1.0$\pm$0.2$\times 10^{21}$ atom cm$^{-1}$ and photon index $\Gamma$ = 1.46$\pm$0.05; $\chi^2$/dof = 211/204. The 0.3--10 keV luminosity was 2.08$\pm$0.08$\times$10$^{38}$ erg s$^{-1}$. Uncertainties are quoted at a 90\% confidence level.  Figure~\ref{1575spec}  shows  the unfolded 0.3--7.0 keV spectrum  multiplied by the channel energy, assuming the best fit  absorbed power law model.

When a disk blackbody component was added to the power law emission, XSPEC set the inner disk temperature to 8.2$\times$10$^{-4}$ keV, with $N_{\rm H}$ = 1.0$\pm$0.2$\times 10^{21}$ atom cm$^{-1}$, and $\Gamma$ = 1.46$\pm$0.05; $\chi^2$/d.o.f = 211/202. Hence there is no trace of a disk component in the 0.3--7.0 keV spectrum. We therefore conclude that RX\thinspace J0042.3+4115 was in its low state during this observation; since  the theoretical upper luminosity limit for low states in neutron star X-ray binaries is $\sim$3$\times 10^{37}$ erg s$^{-1}$ , RX\thinspace J0042.3+4115 is a likely black hole candidate \citep[see e.g.][ and references within]{barnard11b}. 

\subsubsection{XMM-Newton Rev 381}

The 0.3--10 keV EPIC-pn spectrum of RX\thinspace J0042.3+4115 contained 21765 net source photons over $\sim$55 ks of live time, or $\sim$0.4 count s$^{-1}$. The detector was operated in Full Frame mode, with 73.4 ms frame time; hence, pile-up was negligible. 

An absorbed power law model failed to fit the  spectrum; the best fit model yielded $N_{\rm H}$ $\sim$1.5$\times$10$^{21}$ atom cm$^{-2}$ and $\Gamma$ $\sim$1.7, but $\chi^2/$dof = 412/348, with a null hypothesis probability of 0.011. 

We also tried an absorbed disk blackbody model, since this is characteristic of the thermal high state identified in black hole binaries \citep[e.g.][]{mr06}.  The best fit column density was a factor $\sim$3 lower than the  Galactic line-of-sight density (6.7$\times$10$^{20}$ atom cm$^{-2}$), hence we fixed it to this value. This resulted in an inner disk temperature of 1.5 keV, but  $\chi^2$/dof = 765/349 and a null hypothesis probability of 3$\times 10^{-33}$.  Hence, RX\thinspace J0042.3+4115 was clearly not in the thermal high state. 

\begin{figure}
\epsscale{1}
\plotone{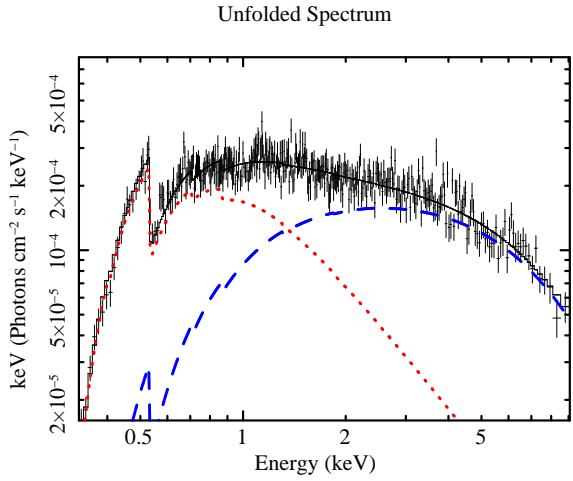}
\caption{Unfolded spectrum for the 60 ks XMM-Newton  observation Rev 381, multiplied by energy. It is well described by  a disk blackbody (dashed) + power law (dotted)  emission model, suffering line-of-sight absorption. $N_{\rm H}$ = 2.4$\pm$0.6$\times 10^{21}$ atom cm$^{-2}$,  k$T_{\rm in}$ = 2.24$\pm$0.15 keV,  and $\Gamma$ = 3.0$\pm$0.6; $\chi^2$/dof = 343/346. Such a spectrum is characteristic of the steep power law  state seen in black hole binaries; the temperature is higher than is generally observed for Galactic black hole binaries, but is consistent with GRS\thinspace 1915+105 \citep{mr06}.}\label{xmmspec}
\end{figure}

A disk blackbody + power law model described the spectrum very well, with $N_{\rm H}$ = 2.4$\pm$0.06$\times 10^{21}$ atom cm$^{-2}$,  inner disk temperature k$T_{\rm in}$ = 2.24$\pm$0.15 keV, $\Gamma$ = 3.0$\pm$0.6, and $\chi^2$/dof = 343/346. Figure~\ref{xmmspec} shows the unfolded spectrum multiplied by channel energy, assuming the best fit model. The 0.3--10 keV luminosity was 2.41$\pm$0.05$\times$10$^{38}$ erg s$^{-1}$, with the power law component contributing $\sim$45\%. Such a spectrum is consistent with the steep power law black hole binary state described by \citet{mr06}. The disc temperature is rather higher than usual, but   consistent with 
 the Galactic black hole binary system GRS\thinspace 1915+105; \citet{mr06} provide a sample spectral fit for GRS\thinspace 1915+105 with k$T_{\rm in}$ = 2.19$\pm$0.04 keV and $\Gamma$ = 3.46$\pm$0.02. 

Some authors have claimed that such a model is unphysical, because the power law component exceeds the thermal component at low energies;  they argue that the observed soft excess is an artifact of the two component model \citep[see e.g.][]{roberts05, goncalves06}. These arguments  assume a compact  corona that can only access photons from the inner disk. However, there is substantial evidence for extended coronae in X-ray binaries at high luminosities; the ingress times of photo-electric absorption dips in high inclination binaries indicate coronae with diameters of $\sim$20,000--700,000 km \citep{church01,cbc04}, while broadened emission lines in Chandra observations of Cygnus X-2 suggest a hot, dense corona of up to $\sim$10$^{5}$ km \citep{schulz09}.
Such coronae would have access to the soft photons from the outer regions of the disc as well as the hot photons from the inner disc.

Indeed, fitting the spectrum with a more physically motivated model ({\sc diskbb + comptt} in XSPEC) yielded the same values for $N_{\rm H}$ and k$T_{\rm in}$ as the disk blackbody + power law model, along with a seed photon temperature of $\sim$0.02 keV, an electron temperature of $\sim$40 keV and an optical depth $\sim$0.2; $\chi^2$/dof = 343/344; the electron temperature was unconstrained, and good fits were obtained for electron temperatures of 100 keV and 300 keV also, typical for the BH low state. Tying the seed photon temperature to  k$T_{\rm in}$ resulted in an unacceptable fit: $\chi^2$/dof $\ge$ 453/346 (null hypothesis probability  $< 1\times 10^{-4}$). These results are entirely consistent with an extended, optically thin corona, and reject a compact corona that only sees the innermost region of the disk.


\section{Discussion}

Although the original evidence for RX\thinspace J0042.3+4115 being a black hole binary was contaminated by artifacts in the XMM-Newton data reduction software, we now have new evidence that it is indeed a black hole candidate.

Our constraint on the B magnitude (M$_{\rm B}$ $>$ 24.8 at the 3$\sigma$ level) is extremely useful for interpreting the system. RX\thinspace J0042.3+4115 is clearly not located in our galaxy, and there is no background galaxy with B $\la$ 28; hence we locate RX\thinspace J0042.3+4115 in M31.  The 3$\sigma$ upper limit to  $M_{\rm B}$ = $-$0.4, meaning that  a low mass donor is most likely.

 We have observed two distinct spectral states from this system, consistent with the low/hard and steep power law black hole states. The low state was observed at a 0.3--10 keV luminosity of 2.08$\pm$0.08$\times 10^{38}$ erg s$^{-1}$, while the steep power law state was observed at 2.41$\pm$0.05$\times 10^{31}$ erg s$^{-1}$. Since transitions from the low state occur at  $L$ $\la $0.1 $L_{\rm Edd}$ in the 0.01--1000 keV band for NS systems \citep{glad07}, and  in the 15--50 keV band for neutron star and black hole systems \citep{tang11}, we suggest that RX\thinspace J0042.3+4115 was near 0.1 $L_{\rm Edd}$ during Chandra observation 1575. Such a system would require a  $\sim$20 M$_{\odot}$ black hole; this is larger than for any Galactic black hole binary, but smaller than the dynamically confirmed black hole in IC10 X-1, which has a best mass estimate of 32.7$\pm$2.6 $M_{\odot}$, and a lower limit of 23.1$\pm$2.1 $M_{\odot}$  \citep{silverman08}.

RX\thinspace J0042.3+4115 has been persistently bright for the last $\sim$12 years; this is quite unlike the transient behaviour of most  Galactic black hole LMXBs. The only known persistently bright black hole LMXB is GRS\thinspace 1915+105, which has remained bright since its discovery in 1992 \citep{mcclintock06}. The X-ray behavior of GRS\thinspace 1915+105 is unmatched by any Galactic black hole binary, and may be explained by the primary spinning in the same direction as the accretion disk  at extreme speeds \citep[$>$98\% of the maximum][]{mcclintock06}. Such prograde spinning allows the last stable orbit to be significantly closer to the black hole than for a non-spinning black hole, resulting in a higher disk luminosity and also a higher k$T_{\rm in}$ \citep{zhang97}. 

The inner disk temperature for RX\thinspace J0042.3+4115 in the steep power law state (2.24$\pm$0.15 keV)  was remarkably similar to that of GRS\thinspace 1915+105 \citep[2.19$\pm$0.04 keV for an example spectrum,][]{mr06}. Hence, the persistent X-ray emission from RX\thinspace J0042.3+4115 may also be due to extreme prograde spin. However, we note that 5 out of the 6 black hole candidates that we have associated with M31 globular clusters are persistent also  \citep{barnard08,barnard09,barnard11b}; such systems are consistent with theoretical predictions for tidal capture of main sequence donor stars \citep{kalogera04}, or ultra-compact systems with degenerate donors \citep{ivanova10}. Therefore, several mechanisms can promote persistently bright black hole binaries.

Finally we note that compact corona models where the seed photon energy is tied to the inner disk temperature were all rejected by the XMM-Newton Rev 381 spectrum of RX\thinspace J0042.3+4115.  However, free fitting of the seed photon energy yielded a good fit at 0.02 keV; this result gives strong support for an extended corona in the intermediate/steep power law state, which is able to access the cooler photons in the outer disc. We drew the same conclusions for the confirmed black hole + Wolf-Rayet binary IC10 X-1 \citep{barnard10}.



\section*{Acknowledgments}
We thank the anonymous referee for thoughtful comments that significantly improved this paper. We also  thank Z. Li for providing the merged ACIS image. This research has made use of data obtained from the Chandra data archive, and software provided by the Chandra X-ray Center (CXC). This work also used an observation from XMM-Newton, an ESA science mission with instruments and contributions directly funded by ESA member states and the US (NASA). Furthermore, this work has used data from the Hubble Legacy Archive.
R.B. is funded by Chandra grant GO9-0100X and HST grant GO-11013. M.R.G. and S.S.M. are both  partially supported by NASA grant NAS8-03060.



{\it Facilities:} \facility{CXO (ACIS)} \facility{HST (ACS)} \facility{XMM-Newton (EPIC-pn)}.




\clearpage



\end{document}